\documentclass[11pt]{article}
\usepackage[utf8]{inputenc}
\usepackage[T2A]{fontenc}
\usepackage[english]{babel}
\usepackage{amsmath,amssymb,amsthm}
\usepackage[x11names, rgb]{xcolor}
\usepackage{fullpage}
\usepackage{verbatim}
\usepackage{tikz}
\usepackage{bbold,xspace}
\usepackage[textwidth=23mm,colorinlistoftodos]{todonotes}
\usetikzlibrary{calc,arrows,shapes,backgrounds,patterns,fit,decorations,decorations.pathmorphing}

\DeclareMathOperator{\poly}{poly}
\DeclareMathOperator{\polylog}{polylog}

\newtheorem{theorem}{Theorem}
\newtheorem{lemma}{Lemma}
\newtheorem{remark}{Remark}
\newtheorem{definition}{Definition}
\newtheorem{corollary}{Corollary}

\newcommand{\citeN}[1]{\cite{#1}}
\newcommand{\newb}[1]{#1}
\newcommand{\newd}[1]{#1}

\renewenvironment{proof}[1][]{\par
  \pushQED{\qed}%
  \normalfont 
  \trivlist
  \item[\hskip\labelsep
        \itshape
    Proof\xspace#1.]\ignorespaces
}{%
  \popQED\endtrivlist
}

\title{Families with infants: speeding up algorithms for NP-hard problems using FFT\thanks{This paper is based on the same results as~\cite{GKM2013b}, but the presentation of the results and the whole discuss have been reworked substantially. Research is partially supported by the Government of the Russian Federation
(grant 14.Z50.31.0030).}}
\author{
Alexander~Golovnev
\thanks{New York University}
\and
Alexander~S.~Kulikov 
\thanks{St.~Petersburg Department of Steklov Institute of Mathematics of the Russian Academy of Sciences}
\and
Ivan~Mihajlin
\thanks{St.~Petersburg Department of Steklov Institute of Mathematics of the Russian Academy of Sciences}
}
\date{}

\begin{document}

\maketitle

\begin{abstract}
Assume that a group of people is going to an excursion and our task is to seat them into buses with several constraints each
saying that a pair of people does not want to see each other in the same bus.
This is a well-known coloring problem and it can be solved in $O^*(2^n)$ time by the 
inclusion-exclusion principle as shown 
by 
Bj\"{o}rklund, Husfeldt, and Koivisto in 2009.
Another approach to solve this problem in $O^*(2^n)$ time is to use the fast Fourier transform. A graph is $k$-colorable if and only if the $k$-th power of a polynomial containing a monomial 
$\prod_{i=1}^n x_i^{[i \in I]}$
for each independent set $I \subseteq [n]$ of the graph, contains the monomial $x_1x_2\ldots x_n$. 

Assume now that we have additional constraints: the group of people
contains several infants and these infants should be accompanied by their relatives in a bus. We show that if the number of infants is linear
then the problem can be solved in $O^*((2-\varepsilon)^n)$ time. 
We use this approach to improve known bounds for several NP-hard problems
(the traveling salesman problem, the graph coloring problem, the problem
of counting perfect matchings) on graphs of bounded average degree,
as well as to simplify the proofs of several known results.
\end{abstract}
\section{Introduction}
In this paper we consider algorithms for three classical hard problems: the traveling salesman problem, the chromatic number problem, and the problem of counting
perfect matchings.
\newb{$O^*(2^n)$ algorithms by \citeN{B1962} and \citeN{HK1962} for the traveling salesman problem have been known for more than 50 years already ($n$~is the number of vertices of an input graph, $O^*$ hides polynomial factors of input length).}
The upper bound $O^*(2^n)$ for the chromatic number problem
is proved by \citeN{BHK2009}. The number of perfect matchings can be computed in time $O^*(2^{n/2})$ as shown by \citeN{B2012} (this matches the bound by \citeN{R1963} for bipartite graphs).

%
%

For all three problems mentioned above (chromatic number, traveling salesman, counting perfect matchings),
improving the known bounds for the general case
is a major open problem in the field of algorithms for NP-hard problems. 
Better upper bounds are known however for various special cases.
For Hamiltonian cycle problem, \citeN{B2010} and \newb{\citeN{BHKK2010b}} proved an $O(1.66^n)$ bound for the symmetric case (i.e., undirected graphs), \citeN{CKN2013} proved an $O^*(1.89^n)$ bound for directed bipartite graphs.
In~\cite{BHKK2008,BHKK2010,CP2013} 
better upper bounds are \newb{proven} for graphs of bounded degree (the three considered problems are known to be NP-hard even on graphs of bounded degree).

We present a new approach to get bounds of the form
$O^*((2-\epsilon)^n)$ in various special cases. Namely we show that
such a bound follows almost immediately if the corresponding partition problem possesses a certain structure. Informally, this structure can be 
described as follows. Assume that a group of people is going to an excursion and our task is to seat them into buses with several constraints each
saying that a pair of people does not want to see each other in the same bus.
This is the coloring problem and it can be solved in $O^*(2^n)$ time using the inclusion-exclusion method as shown by \citeN{BHK2009}. 
Another approach to solve this problem in $O^*(2^n)$ time is to use the fast Fourier transform (FFT). A graph is $k$-colorable if and only if the $k$-th power of a polynomial containing a monomial 
\newb{$\prod_{i=1}^n x_i^{\newd{[i \in I]}}$} 
for each independent set $I \subseteq [n]$ of the graph contains the monomial $x_1x_2\ldots x_n$. This method 
is mentioned by \citeN{CP2010}; previously, \citeN{KW2009} and \newb{\citeN{BHKK2010b}} used multilinear monomial
detection for solving parameterized problems. 
Assume now that we have additional constraints: the group of people
contains several infants and these infants should be accompanied by their relatives in a bus. Roughly, we prove that if the number of infants is linear
then the problem can be solved in $O^*((2-\varepsilon)^n)$ time using FFT.

Using this approach we unify several known results of this kind. An additional advantage of the approach is \newb{the simplicity of using it} as a black box. Namely, all one needs to do is to reveal the corresponding structure of families with infants. This way, some of the known 
upper bounds for the above mentioned problems on graphs of bounded
degree follow just in a few lines. By using additional combinatorial ideas we also prove the following new results.

For the chromatic number problem, \citeN{BHKK2010} presented an algorithm running in time
$O^*((2-\varepsilon(\Delta))^n)$ on graphs of bounded maximum degree~$\Delta=O(1)$. The algorithm is based on Yate's 
algorithm and M\"{o}bios inversion and thus uses exponential space. We extend this result to a wider class 
of bounded average degree graphs. This closes an open problem concerning the existence of such an algorithm stated
by \citeN{CP2013}.

For the traveling salesman problem on graphs of maximum degree~$\Delta=O(1)$, \citeN{BHKK2008} presented an algorithm running in time $O^*((2-\varepsilon(\Delta))^n)$
and exponential space. \citeN{CP2013} extended the result to graphs of bounded average (rather than maximum) degree. Both algorithms are based on dynamic programming and the savings in the running time comes from an observation that in case of bounded degree graphs an algorithm does
not need to go through all possible subsets of vertices (e.g., a disconnected subgraph does not have a Hamiltonian path for sure). It is also because of the dynamic programming technique that both mentioned algorithms use exponential space. We further extend these results presenting an algorithm running in time \newb{
{$O^*(M(2-\varepsilon(d))^n)$} and polynomial space on directed graphs of average degree~$d$ with integral weights bounded by~$M$.}

\citeN{CP2013} developed an algorithm with running time $O^*((2-\varepsilon(d))^{n/2})$ and exponential space for counting perfect matching in graphs of average degree~$d$. We present an algorithm solving this problem in $O^*((2-\varepsilon(d))^{n/2})$ time and polynomial space. Several bounds of this kind are already known for bipartite graphs~\cite{ALS1991,BF2002,SW2005,RBR2009,IW2012,CP2013}.
\section{Notation}
Let $G=(V,E)$ be a simple undirected graph. Throughout the paper we implicitly assume that the set of vertices of a graph under consideration is $V=\{1,2, \dots, n\}$. For simplicity, we consider undirected graphs only (whether a graph is directed or not is only important for the traveling salesman problem; the presented algorithm works for both undirected and directed graphs).

By $d(G)$ and $\Delta(G)$ we denote the average and the maximum degree of $G$ (we omit $G$ if it is clear from the context).
$N_G(v)$ is a \emph{neighborhood} of $v$ in~$G$, i.e., all the neighbors of $v$ in~$G$, and $N_G[v]=N_G(v) \cup \{v\}$ is its \emph{closed neighborhood}.
For $S \subseteq V$, by $G[S]$ we denote a subgraph of $G$ induced by $S$. We use $G\setminus S$ as a shortcut for $G[V \setminus S]$.

The \emph{square} of $G=(V,E)$ is a graph $G^2=(V,E')$ where $E' \supseteq E$ is
\[E'=\{(u,v) \colon \text{ there is a path of length at most $2$ from $u$ to $v$ in $G$}\} \,.\] Note that $\Delta(G^2) \le (\Delta(G))^2$ and hence one can easily find an independent set of size $\frac{n}{(\Delta(G))^2+1}$ 
in~$G^2$.

Following~\cite{CP2013}, by 
$V_{>c}$ we denote a subset of vertices $V$ of degree greater than~$c$.
$V_{<c}$, $V_{=c}$, $V_{\le c}$, $V_{\ge c}$ are defined similarly.
By $\mathbb{Z}_{\ge c}$ we denote the set of all integers greater than or equal to~$c$.

\newb{For an positive integer $k$, by $[k]$ we mean the set of all positive integers less than or equal to $k$. While for a Boolean expression $P$, by $[P]$ we mean $1$ if $P$ is true and $0$ otherwise (this is the standard Iverson bracket notation).}

Throughout the paper by $\varepsilon$ we denote a positive constant that does not depend on the size of a graph.
\section{Toolkit}
This section describes the main toolkit for proving upper bounds for NP-hard problems using FFT. 
In the first two subsections we provide the main such tools without proofs and give short proofs of several recently proved upper bounds using these tools. All the proofs are given in the third subsection.

\subsection{FFT}
In this subsection we remind and adjust the FFT technique for our goals. 
\newb{In the first two theorems we deal with univariate polynomials, and then we proceed to multivariate polynomials.}

\begin{theorem}
\label{fft:exp}
Let $P(x)=\sum_{i=0}^np_ix^i,Q(x)=\sum_{i=0}^nq_ix^i$ be polynomials of degree at most $n$ with non-negative integer coefficients less than $W$. If $P(x)$ and $Q(x)$ are given as lists $(p_0,\ldots, p_n)$ and $(q_0,\ldots, q_n)$ of coefficients then the list of coefficients of their product $S(x)=P(x)\cdot Q(x)$ can be found in time and space $n\polylog(n,W)$.
\end{theorem}

\begin{theorem}
\label{fft:poly}
Let $P(x)=\sum_{i=0}^np_ix^i$ be a polynomial of degree at most $n$ with non-negative integer coefficients less than $W$. Given an arithmetic circuit $C(x,p)$ of size $\polylog(n,W)$ which evaluates $P$ modulo a prime $p=O(n\polylog(n,W))$ at an integer point $x$, any coefficient of $P(x)$ can be found in time $n\polylog(n,W)$ and space $\polylog(n,W)$.
\end{theorem}

We use the previous two theorems to prove the following result dealing with multivariate polynomials.

\begin{theorem}
\label{fft:multi}
Let $P_1(x_1,\ldots,x_n,z),\ldots, P_n(x_1,\ldots,x_n,z)$
be polynomials of $n+1$ variables with non-negative integer coefficients less than $W$, where individual degrees of $x$'s are \newb{at most} $n$, and the degree of $z$ is at most $d$. Let also $\Pi(x_1,\ldots,x_n,z)=\prod_{i=1}^{n}P_i$ be the product of the polynomials. The smallest $k\le nd$, s.t. $\Pi$ contains the monomial $x_1\ldots x_nz^k$, and the coefficient of this monomial in \newb{$\Pi$} can be found
\begin{enumerate}
\item in time $d2^{n}\poly(n,\log{W})$ and space $2^{n}\poly(n,\log(W,d))$, if $P_i$'s are given as lists of monomials and the length of each list is at most $2^n$;
\item in time $d2^{n}\poly(n,\log{W})$ and space $\poly(n,\log(W,d))$, if $P_i$'s are given as circuits $C_i(x,p)$ of size $\poly(n,\log(W,d))$ where each $C_i$ evaluates $P_i$ modulo a prime $p=O(d2^n\poly(n,\log{W}))$ at an integer point $x$.
\end{enumerate}
\end{theorem}

\begin{remark}
In the following, we will use Theorem~\ref{fft:multi} with $k \le n$ polynomials.
For this, it is enough to set $P_{k+1} \equiv P_{k+2} \equiv \ldots \equiv P_{n} \equiv 1$.
\end{remark}

To show the usefulness of Theorem~\ref{fft:multi} we reprove the following recent result by \citeN{LN2010}.

\begin{lemma}\label{lm:tspln}
The traveling salesman problem on graphs with integer weights from $[M]$
can be solved in time $O^*(M2^n)$ and space \newb{$\poly(n, \log M)$.}
\end{lemma}

Before proving this lemma, we state a technical fact that will be used again further in the text.

\begin{lemma}\label{lm:tsppoly}
For a graph $G$, let $P(x_1, \ldots, x_n,z)$ be a polynomial defined as follows:
\begin{equation}\label{eq:tsppoly}
P(x_1, \ldots, x_n) = \sum\limits_{\text{closed walk } 1 \to i_2 \to \dots \to i_n \to 1 \text{ of \newb{weight} $w$}}x_1x_{i_2}\ldots x_{i_n}z^w \, .
\end{equation}
If the edge weights of $G$ are from $[M]$, then \newb{$P \text{ mod } p$ where $p=O^*(M2^n)$} can be evaluated in time
and space $\poly(n, \log M)$ at any input \newb{$x_1, \ldots, x_n,z<p$}.
\end{lemma}

\begin{proof}
The polynomial can be evaluated by the standard dynamic programming.
Namely, let $Q_{i,k}(x_1, \ldots, x_n,z)$ be a polynomial
containing all walks of length $k$ starting at the vertex $1$ and ending at the 
vertex~$i$. The polynomials $Q_{i,k}$
can be evaluated recursively using a straightforward relation for $k>1$:
\[Q_{i,k}(x_1, \ldots, x_n,z)=\sum\limits_{(j,i) \in E(G)}Q_{j,k-1}(x_1, \ldots, x_n)\cdot x_i \cdot z^{w(j,i)} \, .\]
The initial setting is 
\[Q_{i,\newb{0}}(x_1, \ldots, x_n, z)=x_i \, .\]
Then,
\[P(x_1, \ldots, x_n,z)=\sum\limits_{(i,1) \in E(G)}Q_{i,n-1}(x_1, \ldots, x_n)\cdot z^{w(i,1)} \, .\] 
It is easy to see that the degree of $z$ in $P$ is at most $nM$\newb{, 
and that the evaluation of $P \text{ mod } p$ requires only $\poly(n, \log M)$ time and space.}
\end{proof}

\begin{proof}[of Lemma~\ref{lm:tspln}]
All one needs to do is to find the smallest $k$ such that the polynomial $P$ defined by 
(\ref{eq:tsppoly}) contains the monomial $x_1\ldots x_nz^k$. \newb{By} Lemma~\ref{lm:tsppoly}, $P$ can be evaluated in \newb{$\poly(n, \log M)$} time and space.
Now, the statement follows from the second part of Theorem~\ref{fft:multi} with $P_1=P$ and $P_2 \equiv \ldots \equiv P_n \equiv 1$.
\end{proof}

In the next subsection we show also how to use Theorem~\ref{fft:multi} to find
the chromatic number of a graph in time and space $O^*(2^n)$.

\subsection{Partition problems and families with infants}
\begin{definition}[partition problem]
Let $1 \le k \le n$ be integers and let $\mathcal{F}=\{\mathcal{F}_1, \ldots, \mathcal{F}_k\}$ where each $\mathcal{F}_i \subseteq 2^{[n]}$ is a family of subsets of~$[n]$. An~\emph{$(n,k,\mathcal{F})$-partition problem} is to represent $[n]$ as a disjoint union of $k$ sets from~$\mathcal{F}_i$'s:
\([n]=F_1 \sqcup \ldots \sqcup F_k \text{, where } F_i \in \mathcal{F}_i, \forall 1 \le i \le k \, .\)
\end{definition}

This definition is similar to the one used by \citeN{BHK2009}, the only difference being that in the definition above the families $\mathcal{F}_i$'s are not necessarily equal. 
The brute force search algorithm for this problem takes time $O^*(\max_{1 \le i \le k}|\mathcal{F}_i|^k)$.
Using FFT one can easily prove an upper bound $O^*(2^n)$ which beats the previously mentioned bound in many interesting cases.

There exists a natural one-to-one mapping between families of subsets of $[n]$ and zero-one multilinear polynomials of $n$ variables: for $\mathcal{F} \subseteq 2^{[n]}$, denote by $P_{\mathcal{F}}$ the following polynomial:
\newb{\[ P_{\mathcal{F}}(x_1, \ldots, x_n)=\sum\limits_{F \in \mathcal{F}}\prod\limits_{i=1}^{n}x_i^{[i \in F]} \, .\]}
I.e., elements of $\mathcal{F}$ correspond to monomials of $P_{\mathcal{F}}$.
Conversely, for a zero-one multilinear polynomial $P$ \newb{we} denote by $\mathcal{F}_P$
the corresponding family of subsets of $[n]$. There is also a natural extension
of the function $\mathcal{F}_P$ for all polynomials (but not just 0-1 multilinear polynomials): $\mathcal{F}_P$ contains all subsets $S \subseteq [n]$ for which
$P$ contains a monomial whose variable set is exactly $S$ (thus, the coefficient of this monomial must be non-zero and for each $i \in S$ the degree of $x_i$ in this monomial must be at least $1$). For a monomial $m$, by $F(m)$ we denote the corresponding subset of $[n]$ and by $\text{deg}(m)$ we denote the total degree of~$m$. (E.g., for $m=x_2^4x_3x_9^2$, $F(m)=\{2,3,9\}$ and $\text{deg}(m)=7$).

There is a straightforward reduction from the partition problem to the multilinear monomial detection problem.

\begin{corollary}\label{cor:polyvspartition}
There is a solution for an $(n,k,\mathcal{F})$-partition problem if and only if the polynomial $\prod_{i=1}^kP_{\mathcal{F}_i}$ contains the monomial $x_1\ldots x_n$.
\end{corollary}
Combined with the first part of Theorem~\ref{fft:multi} this gives us the following useful corollary.
\begin{corollary}\label{cor:simple}
Let $\mathcal{F}=\{\mathcal{F}_1, \ldots, \mathcal{F}_k\}$ be a family of subsets of $[n]$ given as a list. Then the $(n,k,\mathcal{F})$-partition problem can be solved in $O^*(2^n)$ time and space.
\end{corollary}

Note that Corollary~\ref{cor:simple} immediately implies $O^*(2^n)$ upper bound for such problems as domatic number and chromatic number. These bounds were proved relatively recently by \citeN{BHK2009} using the inclusion-exclusion method.


Below, we formally define a combinatorial structure called families with infants that
allows to 
prove stronger than $O^*(2^n)$ upper bounds.

\begin{definition}[families with infants for subsets]
Let $\mathcal{S} \subseteq 2^{[n]}$.\\ $\mathcal{R}=((R_1,r_1), \dots, (R_p,r_p))$ 
is called a \emph{$(p,q)$-system of families with infants for $\mathcal{S}$} if all of the following conditions are satisfied:
\begin{enumerate}
\item for all $i=1,\dots,p$\newb{:} $r_i \in R_i\subseteq [n]$; $r_i$ is called \emph{an infant} and all the elements of $R_i\setminus \{r_i\}$ are called \emph{relatives} of $r_i$; the sets $R_i$ are called \emph{families};
\item the size of each family $R_i$ is at most $q$;
\item $pq \le n$;
\item all families $R_i$'s are pairwise disjoint: \newb{$R_i\cap R_j=\emptyset$ for $i\neq j$};
\item in any set $S$ of $\mathcal{S}$ each infant is accompanied by at least one of its relatives:
\begin{equation}\label{eq:mainprop}
\text{$\forall S \in \mathcal{S}$, if $r_i \in S$ then $|F \cap R_i| \ge 2$.}
\end{equation}
\end{enumerate}
\end{definition}

\begin{definition}[families with infants for partitions problems]
$\mathcal{R}$ is called a system of families with infants for an $(n,k,\mathcal{F})$-partition problem where $\mathcal{F}=(\mathcal{F}_1, \ldots, \mathcal{F}_k)$ if it
is so for all $\mathcal{F}_i$, $i \in [k]$.
\end{definition}

\begin{definition}[families with infants for polynomials]
$\mathcal{R}$ is called a system of families with infants for a polynomial $P$ over
$n$ variables if it is so for $\mathcal{F}_P$.
\end{definition}

The next theorem \newb{constitutes} the main technical result of the paper saying
that if a problem possesses \newb{the} property of families with infants then one
can solve the corresponding problem in time $O^*((2-\varepsilon)^n)$.
This property is particularly easy to show for problems
on graphs of bounded degree (either maximum or average).
%

\begin{theorem}\label{thm:maininfants}
Let $P_1, \ldots, P_n$ be polynomials of $x_1, \ldots, x_n, z$ 
as in Theorem~\ref{fft:multi} \newb{(the coefficients are from $[W], \deg{x_i}\le n, \deg{z}\le d$)}.
Let also $P'_i(x_1, \ldots, x_n)=P_i(x_1, \ldots, x_n,1)$, 
and let $\mathcal{R}=((R_1,r_1), \dots, (R_p,r_p))$ be a $(p,q)$-system of families with infants for \newb{$P_1', \ldots, P_n'$}. 
Then the smallest $k$ such that $\Pi=\prod_{i=1}^{n}P_i$ contains the monomial $x_1\ldots x_nz^k$
can be found in time
\newb{
\begin{equation}\label{eq:maineq}
O^*\left(d\cdot2^{n} \cdot \left(\frac{2^q-1}{2^q}\right)^p \cdot 2^q\right)
\end{equation}
}
\begin{enumerate}
\item and the same space, if the polynomials $P_i$'s are multilinear in $x_1, \ldots, x_n$ and are given as lists of monomials;
\item and space $\poly(n, \log W)$ if $P_i$'s are given by curcuits of size $\poly(n, \log W)$.
\end{enumerate}
In particular, if $q=O(1)$ and $p=\Omega(n)$ 
then the upper bound on the running time is \newb{$O^*(d(2-\varepsilon)^n)$}, \newb{where $\varepsilon$ does not depend on $n$.}
\end{theorem}

Roughly, the savings in the running time comes from the fact that 
while looking for a valid partition of $[n]$ one can avoid the case $F \cap R_i=\{r_i\}$
(i.e., instead of considering all $2^q$ possibilities of $F \cap R_i$, one considers
$2^q-1$ of them).

%

Theorem~\ref{thm:maininfants} also implies the $O^*((2-\varepsilon)^n)$ upper bound
for a partition problem with a $(\Omega(n), O(1))$-system of families with infants.

\begin{corollary}\label{cor:partitioninfants}
Given $\mathcal{F}=\{\mathcal{F}_1, \ldots, \mathcal{F}_k\}$ where $k \le n$ and $\mathcal{F}_i \subseteq 2^{[n]}$ as a list of subsets and an $(\Omega(n), O(1))$-system of families with infants $\mathcal{R}$ for the $(n,k,\mathcal{F})$-partition problem one can solve this problem in time and space
$O^*((2-\varepsilon)^n)$.
\end{corollary}

As an illustration of using Corollary~\ref{cor:partitioninfants} we replicate a result from~\cite{BHKK2010}. In the (decision version of) domatic number problem the question is \newb{to check whether it is possible} to partition the set of vertices into $k$ dominating sets. 

\begin{lemma}\label{lemma:domatic}
The domatic number problem in a graph of maximum degree $\Delta=O(1)$ can be solved in time and space $O^*((2-\varepsilon(\Delta))^n)$.
\end{lemma}

\begin{proof}
\label{page:domaticproof}
The domatic number problem is an $(n,k,\mathcal{F})$-problem where each $\mathcal{F}_i$ is just the set of all dominating sets of $G$. By definition, for any $v \in [n]$ and any dominating set $U \subseteq [n]$, $N_G[v] \cap U \neq \emptyset$. This gives a straightforward construction of families with infants.

Find greedily an independent set $I \subseteq V$ of size $p=\frac{n}{\Delta^2+2}$ in $G^2$. Assume w.l.o.g. that $I=\{1, \dots, p\}$. For each $1 \le i \le p$,
let $R_i = N_G[i]$. 
At this point we have at least $n-p(\Delta+1) \ge p$ remaining vertices in $V \setminus \cup_{i=1}^{p}R_i$. So, we can extend each $R_i$ by one vertex
and declare this one additional vertex as the infant of $R_i$.

All $R_i$ have size at most $q=\Delta+2=O(1)$, the total number of $R_i$'s is $p=\frac{n}{\Delta^2+2}=\Omega(n)$. Clearly $pq \le n$. The constructed sets satisfy the property~(\ref{eq:mainprop}) by \newb{the following reason.
Each $R_i$ is a proper superset of $N_G[v]$ for some $v \in V$, moreover \newd{no} vertex from $N_G[v]$ is the infant of the family~$R_i$. For any dominating set~$U$ and any vertex~$v$, $U \cap N_G[v] \neq \emptyset$. Thus, any dominating set contains at least one relative of $r_i$ (even if it does not contain $r_i$ itself).
}

The upper bound now follows from Corollary~\ref{cor:partitioninfants}.
\end{proof}

Another example is a $O^*((2-\varepsilon)^n)$ algorithm for the traveling salesman problem for graphs of bounded degree. This result
was given by \citeN{BHKK2008}.

\begin{lemma}
The traveling salesman problem on a graph of maximum degree $\Delta=O(1)$ and with integer weights from $[M]$ can be solved in time and space $O^*(M(2-\varepsilon(\Delta))^n)$.
\end{lemma}

\begin{proof}
It is enough to find a system of families with infants for the polynomial $P$
from (\ref{eq:tsppoly}).
For this, we just construct greedily an independent set $I$ of size 
$p=\frac{n}{\Delta^2+1}$ in $G^2$. Assume that $I=\{1, \dots, p\}$ and let $R_i=N_G[i]$, $r_i=i$. This is clearly a $(p,q)$-system of sets with infants
for $q=\Delta+1$, since in each closed walk in $G$ an infant $i$ must be accompanied
by one of its neighbours. Theorem~\ref{thm:maininfants} then implies an upper bound $O^*((2-\varepsilon(\Delta))^n)$.
\end{proof}


%
%
%
%
\subsection{Proofs}

\begin{proof}[of Theorem~\ref{fft:exp}]
Let $m=2W^2n+1$, $\mathbb{Z}_m$ be the ring of integers modulo $m$. Since each coefficient of $S(x)=P(x)\cdot Q(x)$ is a positive integer less than $m$, it \newb{suffices} to find $\bar{S}(x)=\bar{P}(x)\cdot \bar{Q}(x)$, where $\bar{P}(x),\bar{Q}(x),\bar{S}(x)\in \mathbb{Z}_m[x]$. Note that one can perform arithmetic operations in $\mathbb{Z}_m$ in time $\polylog(m)$. Since $2\in \mathbb{Z}_m^*$, $P(x)$ and $Q(x)$ can be multiplied in time $n\polylog(n)\polylog(m)=n\polylog(n,W)$ (see, e.g., Exercise~17.24 in~\cite{S2009}).
\end{proof}

\begin{proof}[of Theorem~\ref{fft:poly}]
Assume one is to find the coefficient $p_m$ of the monomial $x^m$ in $P(x)=\sum_{i=0}^np_ix^i$.
By the prime number theorem, there exist \newb{at least} $k=\log{W}$ primes in the interval $[n,n\polylog(n,W)]$. Using a deterministic polynomial primality-testing algorithm (e.g., the AKS algorithm~\cite{AKS2004}) one can find primes $q_1,\ldots,q_{k}$ in this interval in time $n\polylog(n,W)$. By the Chinese remainder theorem, it suffices to find $p_m$ modulo $q_1,\ldots,q_{k}$ in the desired time and space.

We show how to find $p_m$ modulo $q=q_i$. Since $q=n\polylog(n,W)$, one can factor $q-1$ in deterministic time $n\polylog(n,W)$ using trivial division. Given the factorization of \newb{$q-1=a_1^{b_1}\ldots a_s^{b_s}$, one finds a primitive root of unity in $\omega\in\mathbb{Z}^*_{q}$ in $n\polylog(n,W)$ steps as follows. For every $t\in\mathbb{Z}^*_{q}$, one checks if $t^\frac{q-1}{a_i}\neq1$ for all $i \in [s]$, this takes $qs\polylog(q)=n\polylog(n,W)$ steps.
Now the} coefficient $p_m$ of $x^m$ in $\mathbb{Z}^*_{q}$ equals 
$$\frac{1}{q-1}\sum_{i=0}^{q-2}\omega^{-im}P(\omega^i).$$
Since $P(x)$ can be evaluated in $\mathbb{Z}^*_{q}$ in $\polylog(n,W)$ time and space, one needs $n\polylog(n,W)$ time and only $\polylog(n,W)$ space to find one coefficient of~$P$.
\end{proof}

For a subset $U \subseteq [n]$, let $b(U) \in \{0,1\}^n$ denote the characteristic
vector of the set $U$ (i.e., $b(U)[i]=1$ if and only if $i \in U$). 
In the analysis below we sometimes identify a bit vector $b(U)$ with a non-negative integer between $0$ and $2^n-1$ that it represents.

For a bit vector $b$, we denote the Hamming weight of $b$, \newb{i.e., the number of $1$'s in $b$}, by $w(b)$. Note the following simple fact: for any two non-negative integers
$a$ and $b$,
\begin{equation}\label{eq:carries}
 w(\operatorname{bin}(a))+w(\operatorname{bin}(b)) \ge w(\operatorname{bin}(a+b)) 
\end{equation}
and the equality holds if and only if there are no carries in $a+b$.


\begin{lemma}
\label{lemma:binary}
Let $T_1(x_1,\ldots,x_n),\ldots, T_k(x_1,\ldots,x_n)$ be polynomials of $n$ variables $x_1,\ldots,x_n$ with non-negative coefficients. Let also \[Q_i(x,y)=\sum_{m \in T_{i}}y^{\deg(m)}x^{b(F(m))} \, .\]

Then $T=\prod_{i=1}^kT_i(x_1, \ldots, x_n)$ contains the monomial $x_1\dots x_n$
if and only if $Q(x,y)=\prod_{i=1}^{k}Q_i(x,y)$ contains the monomial $y^nx^{b([n])}$.
\end{lemma}

\begin{proof}
One direction of this statement is straightforward.
If $T$ contains the (multilinear) monomial $x_1\dots x_n$, then
$T_1, \ldots, T_k$ contain multilinear monomials $m_1, \ldots, m_k$ such that
$m_1\dots m_k=x_1\dots x_n$. Then 
$F(m_1) \sqcup \ldots \sqcup F(m_k)=[n]$ and
\[ \prod_{i=1}^{k}y^{\deg(m_i)}x^{b(F(m_i))}=y^nx^{b([n])} \, .\]

For the reverse direction, assume that $Q=\prod Q_i$ contains
the monomial $y^nx^{b([n])}$. Because of the term $y^n$,
there exist $k$ monomials $m_1 \in T_1, \dots, m_k \in T_k$ such that 
$\deg(m_1)+\ldots+\deg(m_k)=n$. In other words, the total number of $1$'s in 
all characteristic vectors of $F(m_i)$'s at most~$n$. Moreover,
\begin{equation}\label{eq:sum}
b(F(m_1))+\ldots+b(F(m_k))=b([n]) \,.
\end{equation}
From (\ref{eq:carries}), one concludes that the equality  (\ref{eq:sum})
is only possible when \[w(b(F(m_1)))+\dots+w(b(F(m_k)))=n\]
and there are no carries in  (\ref{eq:sum}).
This in turn implies that $\{F(m_1), \ldots, F(m_k)\}$ is a partition of~$[n]$.
\end{proof}

\begin{proof}[of Theorem~\ref{fft:multi}]
Let for $1\le i\le n$, $P_i(x_1,\ldots,x_n,z)=\sum_{j=0}^d{z^jT_{ij}(x_1,\ldots,x_n)}$, where $T_{ij}$'s are multilinear polynomials. Let also 
\[Q_i(x,y,z)=\sum_{j=0}^{d} z^j\sum_{m \in T_{ij}}y^{\deg(m)}x^{b(F(m))} \, .\]
From Lemma~\ref{lemma:binary}, $\Pi=\prod P_i$ contains the monomial  $x_1\ldots x_nz^k$ if and only if $Q=\prod Q_i$ contains the monomial  $z^ky^nx^{b([n])}$.

Now it suffices to show how to efficiently find the sum of coefficients of the monomials $z^i y^n x^{b([n])}$ for $i
\le k$. Indeed, since all the coefficients are positive, the binary search on $k$ gives us the smallest $k$ s.t. $\Pi$ contains $x_1\ldots x_nz^k$, and the coefficient of this monomial in $\Pi$. 

It is easy to see that the degree of $z$ in $Q=\prod_{i=1}^{n}Q_i(x,y,z)$ does not exceed $dn$. Similarly, the degree of $y$ does not exceed $n^2$. Therefore, in order to obtain univariate polynomials we can use Kronecker substitution~\cite{K1882}. Namely, we replace $y$ by $z^{dn+1}$, and $x$ by $z^{(dn+1)(n^2+1)}$. Thus, for each $1 \le i \le n$ we consider a univariate polynomial $Q'_i(z)$:
\[Q'_i(z)=\sum_{j=0}^{d} z^j\sum_{m \in T_{ij}}z^{(dn+1)|m|}z^{(dn+1)(n^2+1)b(F(m))} \, .\]
It it easy to see that the coefficient of $z^{a_1+(dn+1)\cdot a_2+(dn+1)(n^2+1)a_3}$ (where $a_1\le dn, a_2\le n^2$) in $Q'(z)=\prod Q'_i(z)$ equals the coefficient of $z^{a_1}y^{a_2}x^{a_3}$ in $Q=\prod Q_i(x)$.
In other words, we associate an integer from $[0..(dn+1)(n^2+1)2^n]$ with each monomial $z^i m$, where $m$ is a monomial over $\{x_1,\ldots, x_n\}$. This integer is an encoding of $i$ and $F(m)$ in $n+3\log{n}+\log{d}$ bits, s.t. the first $n$ bits indicate elements of $F(m)$, the next $\log{n}$ bits are zeros, then $\log{n}$ bits are the binary expansion of $|m|$, $\log{n}$ zeros again, and the last $\log{d}$ bits encode $i$.
We need to find the sum of coefficients of \[z^{i+n(nd+1)+(nd+1)(n^2+1) b([n])}\] in $Q'(z)=\prod_{i=1}^{k}Q'_i(z)$ for $i\le k$.
 In order to do that, we multiply $Q'$ by $(1+z+\ldots + z^k)=(z^{k+1}-1)/(z-1)$. The coefficient of 
\[z^{k+n(nd+1)+(nd+1)(n^2+1) b([n])}\] in the obtained polynomial $Q''(z)=(z^{k+1}-1)/(z-1) Q'(z)$ is equal to the sum of coefficients of \[z^{i+n(nd+1)+(nd+1)(n^2+1) b([n])}\] in $Q'(z)$ for all $i\le k$.

Note that the degree of $Q''(z)$ is $O(d2^n\poly(n))$. Now Theorem~\ref{fft:exp} and Theorem~\ref{fft:poly} finish the proof. Namely, to prove $(1)$, we just apply the FFT from Theorem~\ref{fft:exp} $(n-1)$ times. In order to prove $(2)$, we note that $Q''(z) \bmod p$ can be easily obtained from the values $P_i(x_1,\ldots,x_n,z) \bmod p$.
\end{proof}

%

\begin{definition}
For a matrix $M=(M[i,j])_{0\le i \le p-1, 0 \le j \le q-1} \in \mathbb{Z}_{\ge 0}^{p \times q}$ let
\begin{eqnarray*}
\operatorname{colweight}(M,j)&=&\sum_{i=0}^{p-1}M[i,j]\,,
\operatorname{weight}(M)=\sum_{i=0}^{p-1}\sum_{j=0}^{q-1}M[i,j]=\sum_{j=0}^{q-1}\operatorname{colweight}(M,j)\,,\\
\operatorname{rowcode}(M,i)&=& -M[i,0]+\sum_{j=1}^{q-1}2^j\cdot M[i,j]\,,
\operatorname{rowsum}(M)=\sum_{i=0}^{p-1}\operatorname{rowcode}(M,i)\,,\\
\operatorname{code}(M)&=&\sum_{i=0}^{p-1}(2^q - 1)^i\cdot \operatorname{rowcode}(M,i)\,.
\end{eqnarray*}
\end{definition}

\begin{definition}
A matrix $M \in \mathbb{Z}_{\ge 0}^{p \times q}$ is called \emph{row-normalized} if $\operatorname{rowcode}(M,i) \ge 0$ for all $0 \le i \le p-1$.
\end{definition}
\begin{remark}
In the analysis below we will need the following simple estimates. Let $E \in \{0,1\}^{p \times q}$. Then
\begin{enumerate}
\item $\operatorname{rowcode}(E,i)<0$ if and only if $E[i]=[1,0,0,\ldots,0],$
\item $\operatorname{rowcode}(E,i) \le 2^q-2 $.
\item $\operatorname{code}(E) \le (2^q-2)\cdot \sum_{i=0}^{p-1}(2^q-1)^i<(2^q-1)^p$ (assuming $q \ge 2$).
\end{enumerate}
\end{remark}

The following fact is well known so we state it without a proof.

\begin{lemma}
\label{lemma:system}
The expansion of $X\in\mathbb{Z}_{\ge 0}$ in powers of $b>1$ as $X=\sum_{i=0}^{\infty}x_i\cdot b^i, x_i\ge0$ has the minimal value of the sum of digits $\sum_{i=0}^{\infty}x_i$ if and only if $\forall i: 0\le x_i<b$ (i.e., $X$ is written in the numeral system of base $b$).
\end{lemma}

\begin{lemma}
\label{lemma:coding}
Let $q\ge 2$ and $E \in \{0,1\}^{p \times q}$
and $M \in \mathbb{Z}_{\ge 0}^{p \times q}$ be row-normalized matrices.
If
$\operatorname{colweight}(M,0)=\operatorname{colweight}(E,0)$, 
$\operatorname{weight}(M)=\operatorname{weight}(E)$,
$\operatorname{rowsum}(M)=\operatorname{rowsum}(E)$,
$\operatorname{code}(M)=\operatorname{code}(E)$,
then $M=E$.
\end{lemma}

\begin{proof}
The claim follows from Lemma~\ref{lemma:system}.
Since for all $i$, $\operatorname{rowcode}(E,i) \le 2^q-2 $, $\operatorname{code}(E)$ has the minimal sum of digits in base $(2^q-1)$ system. This in turn implies that for each $i$, $\operatorname{rowcode} (E,i)=\operatorname{rowcode}(M,i)$. Then the first columns of matrices $E$ and $M$ are equal modulo $2$, because parities of rowcodes depend only on the first column. Since $\operatorname{colweight}(M,0)=\operatorname{colweight}(E,0)$ we conclude that the first columns of $M$ and $E$ are equal.

Now each $\operatorname{rowcode}(E,i)$ has the minimal sum of digits in the system of base $2$, which means that $\operatorname{weight}(E)$ has the minimal possible value for these $\operatorname{rowcodes}$. It follows from Lemma~\ref{lemma:system} that each $M[i,j]$ must be equal to $E[i,j]$.
\end{proof}

\begin{definition} 
An injective function $\alpha \colon U \to \{0,\dots,p-1\} \times \{0,\dots, q-1\}$ is called a \emph{matrix representation} of a set $U$. For such $\alpha$ and $S \subseteq U$, a~\emph{characteristic matrix} $M_\alpha(S) \in \{0,1\}^{p \times q}$ is defined as follows: $i \in U$ if and only if $M_\alpha(S)[\alpha(i)]=1$.
\end{definition}

\begin{proof}[of Theorem~\ref{thm:maininfants}]
Let $\mathcal{R}=((R_1,r_1), \dots, (R_p,r_p))$ be a {$(p,q)$-system of families with infants}
for a \newb{$P_1', \ldots, P_n'$}. Append arbitrary elements from $[n]$ to families so that the size of each family equals $q$ and the families are still disjoint (this is possible since $pq \le n$). Denote the union of  families by $R$ and the rest of $[n]$ by~$L$. 
For each family~$R_i$, fix an order of its elements such that
the $0$th element is~$r_i$. Now consider a matrix representation $\alpha \colon [n] \to \{0,\dots,p-1\} \times \{0,\dots, q-1\}$ defined as follows.  If $v$ is the $j$th element of $R_i$, then $\alpha(v)=(i,j)$.
We encode each monomial $m_i \in P_i$ by parts. We encode elements from $F(m_I)\cap L$ using the standard technique from Lemma~\ref{lemma:binary}. To encode elements from $F(m_i)\cap R$ we use the characteristic matrix $M_\alpha(F(m_i)\cap R)$. Note that $M_\alpha(F(m_i)\cap R)$ is a {row-normalized matrix}, because if $F(m_i)$ contains an infant $r_i$ of a family~$R_i$, then it must contain at least one other element from the same row.
Consider the following polynomials for $1 \le i \le k$: $Q_i(u_1,u_2,u_3,u_4,u_5,u_6,z)$
is equal to
\[\sum_{m=z^tm_1 \in P_i,F = F(m_1)}u_1^{|F\cap L|}\cdot u_2^{b(F\cap L)} \cdot u_3^{\operatorname{colweight}(M,0)} \cdot u_4^{\operatorname{weight}(M)} \cdot u_5^{\operatorname{rowsum}(M)} \cdot u_6^{\operatorname{code}(M)} \cdot z^t\, ,\]
where $M=M_\alpha(F\cap R)$ and $b(F\cap L)$ is an integer from $0$ to $2^{|L|}-1$.

We claim that $\prod_{i=1}^nP_i$ contains the monomial $x_1\ldots x_nz^t$
if and only if $\prod_{i=1}^nQ_i$ contains the monomial
\[u_1^{|L|}u_2^{b(L)}u_3^{\operatorname{colweight}(R,0)} u_4^{\operatorname{weight}(R)} u_5^{\operatorname{rowsum}(R)} u_6^{\operatorname{code}(R)} z^t\, .\] 
Indeed, as it was shown in Lemma~\ref{lemma:binary}, $u_1^{|L|}u_2^{b(L)}$ corresponds to partitions of $L$. Lemma~\ref{lemma:coding} implies that only partitions of $R$ may have the term 
\[u_3^{\operatorname{colweight}(R,0)} u_4^{\operatorname{weight}(R)} u_5^{\operatorname{rowsum}(R)} u_6^{\operatorname{code}(R)}\,.\] Note that the degrees of $u_1,u_3,u_4$ in $\prod_{i=1}^{k}Q_i(u_1,u_2,u_3,u_4,u_5,u_6,z)$ are bounded from above by $n^2$, the degree of $u_2$ is bounded by $n\cdot2^{|L|}$, the degree of $u_5$ is bounded by $n^2\cdot2^q$, the degree of $u_6$ is bounded by $n\cdot(2^q-1)^p$, \newb{the degree of $z$ is bounded by $dn$}.  Now we can apply Kronecker substitution as in the proof of Theorem~\ref{fft:multi}:

\newb{\begin{align*}
u_1&=u\\
u_2&=u^{(n+1)^2}\\
u_3&=u^{(n+1)^32^{|L|}}\\
u_4&=u^{(n+1)^52^{|L|}}\\
u_5&=u^{(n+1)^72^{|L|}}\\
u_6&=u^{(n+1)^92^{|L|}2^q}\\
z&=u^{(n+1)^{10}2^{|L|}2^q(2^q-1)^p}
\end{align*}}

The running time of FFT is bounded by the degree of the resulting univariate polynomial, i.e.
\newb{
\[O^*(\poly(n)d2^{|L|}2^q(2^q-1)^p)=O^*(d2^{n-pq}(2^q-1)^p2^q)=O^*\left(d\cdot2^n\cdot\left(\frac{2^q-1}{2^q}\right)^p \cdot 2^q\right)\,.\]
}
The required statements now follow from Theorems \ref{fft:exp} and \ref{fft:poly}.
\end{proof}

\section{Properties of bounded degree graphs}\label{sec:bounded}
The following three lemmas are \newb{proven} by Cygan and Pilipczuk~\cite{CP2013}. We slightly extend the statements and provide the proofs for the sake of completeness.
 
The lemma below allows to find in a graph a set of vertices of high degree with a better upper bound on its size than given by the standard averaging argument.

\begin{lemma}[\cite{CP2013}, Lemma~3.2]\label{lemma:coresize}
For any graph $G=(V,E)$ of average degree at most $d$, any integer $m\ge 1$ and any $\alpha \ge 1$ there exists 
$m\le D\le M$ such that $ |V_{>D}| \le \frac{nd}{\alpha D}$
where $M=\lfloor me^{\alpha+1}+1\rfloor$.
\end{lemma}
\begin{proof}\label{page:averageproofs}
Clearly,
\[\sum_{i=0}^{\infty}|V_{>i}|=\sum_{i=0}^{\infty}i|V_{=i}| \le nd \,.\]
If, on the other hand, $|V_{>i}|>\frac{nd}{\alpha i}$ for all $m\le i\le M$ then
\[\sum_{i=0}^{\infty}|V_{>i}| \ge \sum_{i=m}^{M}|V_{>i}|>\frac{nd}{\alpha}\sum_{i=m}^{M}\frac{1}{i}=
\frac{nd}{\alpha}\left(\sum_{i=1}^{M}\frac{1}{i}-\sum_{i=1}^{m-1}\frac{1}{i}\right) \ge
\frac{nd}{\alpha}\left(\ln M-\ln(em)\right) \ge 
nd \, ,\]
where the next to last inequality uses the well-known estimate for the harmonic series:
\[ \ln(i+1) \le 1+\frac{1}{2}+\ldots +\frac{1}{i} \le \ln i + 1=\ln(ie)\,. \]
\end{proof}

\begin{remark}
Such $D$ can be easily found in polynomial time by going through all the values $D=m,\ldots, M$.
\end{remark}

The next lemma shows that one can find 
a subset of vertices of linear size that is independent in the square of a graph.

\begin{lemma}[\cite{CP2013}, Lemma~3.1]\label{lemma:indsetsize}
For any graph $G=(V,E)$ of average degree at most $d$ and maximum degree at most $\Delta$
one can find in polynomial time a subset of vertices $B \subseteq V$ such that
\begin{enumerate}
\item the size of $B$ is linear \newb{ in the number of vertices}: $|B| \ge \frac{n}{6\Delta d}$;
\item degrees of vertices from $B$ are small: for any $v \in B$, $\operatorname{deg}_G(v) \le 2d$;
\item $B$ is independent in $G^2$: for any $u \neq v \in B$, $N_G[u] \cap N_G[v] = \emptyset$.
\end{enumerate}
\end{lemma}
\begin{proof}
Clearly, $|V_{\le 2d}| \ge \frac{nd}{2d}=\frac{n}{2}$. The required set $B$ can be constructed by a straightforward greedy algorithm: while $V_{\le 2d}$ is not empty, take any $v \in V_{\le 2d}$, add it to $B$, and remove from $V_{\le 2d}$ the vertex $v$ together with all its neighbors in~$G^2$.
The number of such neighbors is at most $2d+2d(\Delta-1)=2d\Delta$. Hence 
at each iteration at most $2d\Delta+1$ vertices are removed and the total number of iterations is at least 
\[\frac{|V_{\le 2d}|}{2d\Delta+1} \ge \frac{n}{4d\Delta+2} \ge \frac{n}{6d\Delta} \,.\] 
\end{proof}


The following lemma allows us to find efficiently in a graph $G$ of average degree $d=O(1)$
a subset of vertices $Y$ of high degree such that $(G\setminus Y)^2$ contains a large enough 
independent set. The last inequality in the statement
can be seen as exponential savings in the running time.

\begin{lemma}[\cite{CP2013}, Lemma~3.4]\label{lemma:ugly}
For any constants $\nu \ge 1, \mu <1, a \ge 0, 0< c < 1$ there exists $\beta>0$ such that for any graph $G=(V,E)$ of average degree~$d=O(1)$ one can find in polynomial time subsets 
$A,Y \subseteq V$ such that:
\begin{enumerate}
\item $A \cap Y = \emptyset$;
\item $A$ is an independent set in $(G\setminus Y)^2$;
\item $2|Y|\le |A| \le cn$;
\item each vertex from $A$ has at most $2d$ neighbors in $G\setminus Y$:
$\forall v \in A,\, |\{u \in V \setminus Y \colon (u,v) \in E\}| \le 2d\,;$
\item 
\begin{equation}\label{eq:ugly1}
{|A| \choose |Y|}^a\nu^{|Y|}\mu^{|A|} < 2^{-\beta n}\,.
\end{equation}
\end{enumerate} 
\end{lemma}
\begin{proof}
Let $\alpha=\alpha(d, \nu, \mu, a,c)$ be a large enough constant to be defined later.
Using Lemma~\ref{lemma:coresize} we can find 
\begin{equation}\label{eq:cd}
 \frac{1}{12dc} \le D \le \frac{1}{12dc}{e^{\alpha+1}+1},
\end{equation}
such that $|V_{>D}| < \frac{nd}{\alpha D}$. Let $Y=V_{>D}$. Note that the graph $G\setminus Y$ has average degree at most $d$ and maximum degree at most~$D$. Lemma~\ref{lemma:indsetsize} allows us to find
a subset $A \subseteq V \setminus Y$ such that $A$ is independent in $(G \setminus Y)^2$,
for all $v \in A$, $\operatorname{deg}_{G \setminus Y}(v) \le 2d$ and 
\[|A| \ge \frac{n-|Y|}{6dD} \ge \frac{n}{12dD} \, ,\]
where the last inequality is true when $\alpha \ge 2d$, i.e., $\alpha$ is large enough.
Remove from $A$ arbitrary vertices such that $|A|=\frac{n}{12dD}$.
Because of~(\ref{eq:cd}), $\frac{n}{12dD} \le nc$.
To guarantee that $|A|>2|Y|$ it is enough to take $\alpha \ge 24d^2$.

We now show how to choose $\alpha$ such that the last inequality from the statement is satisfied. Using the well known estimates ${n \choose k} \le \left(\frac{en}{k}\right)^k$
and ${n \choose k} \le {n \choose k'}$ for $k \le k' \le \frac n2$ we get
\[ {|A| \choose |Y|}^a \le \left(\frac{en\alpha D}{12dDnd}\right)^\frac{nda}{\alpha D}=\left(\frac{e\alpha}{12d^2}\right)^\frac{nda}{\alpha D}=(\gamma \alpha)^\frac{nda}{\alpha D} \, ,\]
where $\gamma=\frac{e}{12d^2}$ is a constant. Thus we can upper-bound
(\ref{eq:ugly1}) as follows:
\begin{equation}\label{eq:ugly2}
\left(\left(\gamma \alpha\right)^\frac{da}{\alpha}\left(\nu^d \right)^\frac{1}{\alpha}\mu^\frac{1}{12d}\right)^\frac{n}{D} \, .
\end{equation}
Recall now that $\mu<1$ and note that $(\gamma\alpha)^\frac{da}{\alpha} \to 1$ and
$\left(\nu^d\right)^\frac{1}{\alpha} \to 1$ with $\alpha \to +\infty$. Thus for a large enough $\alpha$, (\ref{eq:ugly2}) is $\left(2^{-\beta'}\right)^\frac{n}{D}$ for a constant $\beta'>0$. It remains to recall that $D < e^\alpha$ and take $\beta=\beta'e^{-\alpha}$.
\end{proof}

\section{The chromatic number problem}
\begin{definition}
In the \emph{list coloring problem} each vertex $v$ of the input graph is assigned a list $L_v$ of allowed colors and the task is to properly color a graph such that each vertex
is given a color from its list.
\end{definition}

To reduce the search space in the list coloring problem we introduce the following problem.

\begin{definition}\label{def:pref}
In the \emph{coloring with preferences problem} besides the lists for all the vertices
of a graph $G$ one is given a set $P$ of pairs $\{(u_1,v_1), \ldots, (u_t,v_t)\}$ of vertices such that 
\begin{enumerate}
\item all the vertices from $P$ are different: $u_i \neq v_j$ for all $1 \le i,j \le t$;
\item the set $\{u_1, v_1, \ldots, u_t,v_t\}$ is independent in $G^2$;
\item $L_{u_i}=L_{v_i}$ for all $1 \le i \le t$.
\end{enumerate} 
The goal is to color the graph properly using an allowed color for each vertex such that
for all $1 \le i \le t$ at least one of the vertices from $N[u_i]$ is assigned the same color as $v_i$.
\end{definition}

\begin{lemma}
Let $G$ be a graph, $\{L_v\}_{v \in V}$ be a set of list colors for its vertices, and
$P=\{(u_1,v_1), \ldots, (u_t,v_t)\}$ be a set as in Definition~\ref{def:pref}. Then 
there is a solution for an instance $(G, \{L_v\})$ of the list coloring problem if and only if
there is a solution for an instance $(G, \{L_v\}, P)$ of the coloring with preferences problem.
\end{lemma}
\begin{proof}
Obviously, if there is a coloring satisfying preferences then it is also a list-coloring.
For the reverse direction, consider a proper list coloring such that for some vertex 
$u_i$ neither of the vertices from $N[u_i]$ is given the same color as $v_i$. One can then change the color of $u_i$ to the one of $v_i$ (this is allowed since $L_{u_i}=L_{v_i}$). This clearly does not violate any coloring constraints and it strictly increases the number of pairs $(u_i,v_i)$ 
satisfying its preference constraints. 
It remains to note that this recoloring cannot \newb{violate} any other preference constraint
since the set of vertices from $P$ is independent in $G^2$.
\end{proof}

Checking whether a graph $G$ has a proper $k$-coloring is a $(V,\mathcal{F},k)$-partition problem where $\mathcal{F}=\{\mathcal{F}_1,\ldots,\mathcal{F}_k\}$, for all $i=1,\ldots,k$, $\mathcal{F}_i=IS(G)$, the set of all independent sets of $G$. Note that Theorem~\ref{thm:maininfants} already implies 
a $O^*(2^n)$ time and space algorithm for the Chromatic Number problem. An algorithm with the same time and space bounds was given recently by \citeN{BHK2009}.

For $k$-coloring with preferences the families $\mathcal{F}_i$ are defined slightly differently. Namely, $\mathcal{F}_i$
consists of all independent sets $I$ of $G$ that can be assigned the same color
without violating any list constraints and preferred color constraints:
\begin{itemize}
\item (list constraints): $\forall v \in I$, $ i \in L_v$;
\item (preference constraints): for all $1 \le i \le k$ such that $v_i \in I$, 
$N_G[u_i] \cap I \neq \emptyset$.
\end{itemize}
Using this interpretation of the coloring with preferences problem,
we give an algorithm solving the Chromatic Number problem
on graphs of bounded average degree in time $O^*((2-\varepsilon)^n)$ and exponential space.

\begin{theorem}
There is an algorithm checking whether for a given graph $G$ of average degree $d=O(1)$ there exists a proper $k$-coloring in time $O^*((2-\varepsilon(d))^n)$
and exponential space.
\end{theorem}
\begin{proof}
First consider the case $k \ge 2d$. Note that $|V_{\ge k}| \le \frac{nd}{k} \le n/2$.
Then one can find a proper $k$-coloring 
of \newb{the} graph $G[V_{\ge k}]$ in time $O^*(2^{n/2})$. Such a coloring can be easily extended to the whole graph (since there always exists a vacant color for a vertex of degree at most $k-1$). Thus, in the following we assume that $k < 2d=O(1)$.

Let $\nu \ge 1, \mu<1,a \ge 0, 0<c<1$ be constants to be defined later and let $A,Y \subseteq V$ be as provided by Lemma~\ref{lemma:ugly}. For $Y$ we try all possible colorings in time $k^{|Y|}$. A~fixed coloring of $Y$ produces a list coloring problem for~$G \setminus Y$.
Let $L$ be one of the most frequent color lists of vertices from~$A$.
Let 
$C=\{v \in A \colon L_v = L\}$. 
Since there are at most $2^k$ different lists,
$|C| \ge |A|/2^k$. 
If $|C|$ is not divisible by $2$ remove an arbitrary vertex from $C$
and let $C=\{u_1, v_1, \ldots, u_t,v_t\}$. Let now $P=\{(u_1, v_1), \ldots, (u_t,v_t)\}$
be a set of preferences.

It is straightforward to find a $(p,q)$-system of families with infants for the resulting 
partition problem. Let the number of families $p$ be equal to $t$. The infant $r_i$ of the $i$-th family is $v_i$ and its family $R_i$ is $N[u_i]\cup\{v_i\}$. This is a system of families
with infants by definition (by choosing a small enough constant $c$ one can guarantee that $pq \le n$).

To estimate the running time, we first recall that
$p=\frac{|C|}{2}=\frac{|A|}{2^{k+1}}$ and $q \le 2d+1$.
Corollary~\ref{cor:partitioninfants} implies that the running time of the resulting algorithm is at most
\[k^{|Y|}\cdot \left( 2^{|V \setminus Y|} \cdot 
\left(\frac{2^q-1}{2^q}\right)^p\right) \le 2^n \cdot \left( k^{|Y|} \cdot 
\left(\frac{2^q-1}{2^q}\right)^p\right) \, .\]

We now choose the constants $\nu,\mu,a$ so that (\ref{eq:ugly1}) implies that the expression in parentheses is at most $2^{-\beta n}$ for a constant $\beta > 0$. Let $a=0$, $\nu=k$, and 
\[\mu=\left(\frac{2^q-1}{2^q}\right)^{\frac{1}{2^{k+1}}}\]
(recall that $k<2d=O(1)$).
It is easy to see that (\ref{eq:ugly1}) then implies that the total running time is $2^{(1-\beta)n}$ for a constant $\beta>0$.
\end{proof}


\section{The traveling salesman problem}

\begin{theorem}\label{thm:tsp}
The traveling salesman problem on graphs of average degree $d=O(1)$ with integer weights from $[M]$ can be solved 
in time \newb{$O^*(M \cdot (2-\varepsilon(d))^n)$} and polynomial space \newb{$\poly(n,\log M)$}.
\end{theorem}
\begin{proof}
%
%
We construct \newb{a system of families with infants} for the polynomial $P$ given by (\ref{eq:tsppoly}).
Let $\mu<1, 0<c<1$ be constants to be defined later. Let $A,Y \subseteq V$ be provided by Lemma~\ref{lemma:ugly}.
Consider an optimal weight Hamiltonian cycle $C$ in the graph. 
Let $Y' \subseteq V$ be the successors of the vertices from $Y$
in the cycle~$C$. We guess the set $A \cap Y'$ 
($O^*\left({|A| \choose |Y|}\right)$ choices)
and let $A'=A \setminus Y'$. Note that $|A'| \ge |A|/2$ (since $|A| \ge 2|Y|$)
and for each vertex $v \in A'$ its predecessor $u$ in the cycle $C$ belongs to $V \setminus Y$.

Let $A'=\{1, \ldots, p\}$. Then for all $i=1,\ldots, p$, $R_i=N_{G\setminus Y}[i]$ and $r_i=i$. Clearly, $|R_i| \le q=2d+1$. By choosing $c<\frac{1}{2d+1}$ we can guarantee that $pq \le n$. All $R_i$'s are disjoint since $A'$ is an independent set in $(G \setminus Y)^2$. Finally, if the set $Y'$ is guessed correctly (i.e., $Y'$ are indeed successors of $Y$ in the optimal cycle~$C$) then $((R_1,r_1), \dots, (R_p,r_p))$ is a $(p,q)$-system of families with infants for $\mathcal{F}_P$. Indeed, if $r_i \in F$ for some closed walk $F \in \mathcal{F}_P$ then two $r_i$'s neighbors in $C$ must lie in $V\setminus Y$, i.e., in $R_i$.

By Theorem~\ref{thm:maininfants} the total running time does not exceed
\[{|A| \choose |Y|} \cdot 2^n \cdot \left(\frac{2^q-1}{2^q}\right)^p\cdot M.\]

Recall that $p=|A'|\ge |A|/2$. Choose $\mu=\left( \frac{2^{2d+1}-1}{2^{2d+1}}\right)^{1/2}$. Then (\ref{eq:ugly1}) implies that 
\({|A| \choose |Y|}\cdot \left(\frac{2^q-1}{2^q}\right)^p < 2^{-\beta n}\)
for a constant $\beta > 0$. Thus the total running time is $O^*(M\cdot (2-\varepsilon)^n)$.
\end{proof}

\section{Counting perfect matchings}
The algorithm for counting perfect matchings shares some common ideas with the algorithm for the traveling salesman problem presented in Theorem~\ref{thm:tsp}.

\begin{lemma}\label{lm:antimatching}
Let $G$ be a graph of average degree $d=O(1)$. Then $\bar{G}$ (the complement of~$G$) contains a matching consisting of $\frac{n}{2}-O(1)$
edges.
\end{lemma}
\begin{proof}
Clearly $|V_{\ge \frac{n}{3}}| \le \frac{nd}{n/3}=3d$. After removing all these vertices from $G$ we get a graph with at least $n-3d$ vertices such that the degree of each vertex is at most $\frac{n}{3}$. This implies that the degree of each vertex in the complement of this graph is at least $n-3d-1-\frac{n}{3}$. This is at least $\frac{n}{2}$ for large enough $n$. By Dirac's theorem~\cite{D1952} this graph is Hamiltonian and hence contains a perfect matching.
\end{proof}

\begin{theorem}
The number of perfect matchings in a graph $G$ with $2n$ vertices of average degree $d=O(1)$ can be found in time $O^*((2-\varepsilon(d))^n)$ and polynomial space.
\end{theorem}
\begin{proof}
Assume that the vertices of $V=\{1, \ldots,2n\}$ are numbered in such a way
that 
\begin{equation}\label{eq:loops}
(1,n+1),(2,n+2), \ldots, (m,n+m) \not \in E
\end{equation}
where $m=n-O(1)$. Such a numbering exists due to Lemma~\ref{lm:antimatching} (and can be efficiently found since we can find a maximum matching in $\bar{G}$ in polynomial time). Following~\citeN{B2012} we reduce the problem of counting perfect matchings to a problem of counting cycle covers of a special type. Construct an auxiliary multigraph $G'=(V',E')$ where $V'=\{1,\ldots,n\}$
and each edge $(i,j) \in E$ is transformed into an edge 
$e=((i \bmod n)+1, (j \bmod n)+1) \in E'$
with the label $l(e)=\{i,j\}$. In other words, we contract each pair of vertices
$(1, n+1), \ldots, (n,2n)$ and on each edge we keep a label showing where it originates from. 
Any two vertices in $G'$ are joined by at most $4$ edges.
The average degree of $G'$ is at most~$2d$. 

Recall that a cycle cover of a multigraph is a collection of cycles such that each vertex belongs to exactly one cycle. In other words, this is a subset of edges such that each vertex is adjacent to exactly two of these edges (and a self-loop is thought to be adjacent to its vertex twice).

An important property of the graph $G'$ is the following:
each perfect matching in $G$ corresponds to a cycle cover $C \subseteq E'$ in $G'$
such that $\cup_{e \in C}l(e)=V$ and vice versa. Indeed, each vertex $i$ in $G'$ is adjacent to exactly two edges. These two edges have different labels so they correspond to edges in the original graph $G$ that match both $i$ and $i+n$. 

We have reduced the problem to counting cycle covers with disjoint labels in~$G'$ (the reduction is due to \citeN{B2012}).
We further reduce the problem to counting cycle covers without self-loops.
Note that by (\ref{eq:loops}), $G'$ has at most $s=O(1)$ self-loops. For each such loop $e=(i,i)$ we can consider two cases: to count the number of cycle covers with $e$ we 
count the number of cycle covers in $G'$ without the vertex~$i$; to count the number 
of cycle covers without $e$ we can just remove the loop $e$ from $G'$ and count the number
of cycle covers. This way, we reduce the problem to $2^s=O(1)$ problems of counting cycle
covers in a multigraph without self-loops.

We will count the number of cycle covers with exactly $t$ cycles for each $t=1,\ldots,n$ separately. For this, we define a polynomial $P(x_1, \ldots, x_n)$ containing a monomial for each closed walk with disjoint labels on adjacent edges (excluding the empty cycle) and compute the coefficient of the monomial $x_1\dots x_n$ in $P^t$. This coefficient divided by $t!$ is exactly the number of cycle covers with $t$ cycles.

To evaluate $P$ efficiently we use dynamic programming again. Let $Q_k(x_1, \ldots, x_n)$ be equal to the sum of all monomials corresponding to all closed walks of length $k$ with 
disjoint labels. Let also $T_{u,v,l_0,l,k}(x_1, \ldots, x_n)$ be equal to the sum of all monomials corresponding to all walks with disjoint labels of length $k$ from $u$ to $v$ such that the label
of the first edge is $l_0$ and the label of the last edge is $l$. 
Note that we only need to ensure that the labels on adjacent edges are disjoint.
Namely, one of the edges adjacent to a vertex $i$ in $G'$ must contain $i$ in its label while the other one must contain $i+n$.
Then
\begin{eqnarray*}
T_{u,v,l_0,l_1,k}(x_1, \ldots, x_n) = \sum\limits_{l_2, e=(w,v) \in E' \colon l_1=l(e), l_1 \cap l_2 = \emptyset} T_{u,w,l_0,l_2,k-1}(x_1, \ldots, x_n) \cdot  x_u \\
Q_k(x_1, \ldots, x_n) =\sum\limits_{e=(u,v) \in E'} \sum\limits_{l_1 \colon l_1 \cap l(e) = \emptyset}\sum\limits_{l_2 \colon l_2 \cap l(e) = \emptyset}T_{v,l_1,u,l_2,k}(x_1, \ldots, x_n)
\end{eqnarray*}

The system of families with infants for $\mathcal{F}_P$ is constructed in exactly the same way like in Theorem~\ref{thm:tsp}.
\end{proof}

\bibliographystyle{plain}
\bibliography{allrefs}

\end{document}